\documentclass[%
 reprint,
superscriptaddress,
 amsmath,amssymb,
 aps,prl,
]{revtex4-2}
\usepackage{hyperref}
\hypersetup{
    colorlinks=true,
    linkcolor=blue,
    filecolor=blue,      
    urlcolor=blue,
    citecolor=blue
    }
\usepackage{braket}
\usepackage{xcolor}
\usepackage{xfrac}
\usepackage{graphicx}
\usepackage{dcolumn}
\usepackage{dsfont}
\usepackage{verbatim}

\usepackage{bm}

\begin{document}

\preprint{APS/123-QED}

\title{Quasiperiodic Floquet-Gibbs states in Rydberg atomic systems}

\author{Wilson S. Martins}

\email{wilson.santana-martins@uni-tuebingen.de}

\affiliation{%
 Institut f{\"u}r Theoretische Physik and Center for Integrated Quantum Science and Technology, \\
 Universit{\"a}t T{\"u}bingen,
 Auf der Morgenstelle 14,
 72076 T{\"u}bingen,
 Germany
}%

\author{Federico Carollo}

\affiliation{Centre for Fluid and Complex Systems, Coventry University, Coventry, CV1 2TT, United Kingdom}

\author{Kay Brandner}%

\affiliation{School of Physics and Astronomy, University of Nottingham, Nottingham, NG7 2RD, UK}
\affiliation{Centre for the Mathematics and Theoretical Physics of Quantum Non-Equilibrium Systems,
University of Nottingham, Nottingham, NG7 2RD, UK}

\author{Igor Lesanovsky}%

\affiliation{%
 Institut f{\"u}r Theoretische Physik and Center for Integrated Quantum Science and Technology, \\
 Universit{\"a}t T{\"u}bingen,
 Auf der Morgenstelle 14,
 72076 T{\"u}bingen,
 Germany
}%
\affiliation{School of Physics and Astronomy, University of Nottingham, Nottingham, NG7 2RD, UK}
\affiliation{Centre for the Mathematics and Theoretical Physics of Quantum Non-Equilibrium Systems,
University of Nottingham, Nottingham, NG7 2RD, UK}

\date{\today}

\begin{abstract}
Open systems that are weakly coupled to a thermal environment and driven by fast, periodically oscillating fields are commonly assumed to approach an equilibrium-like steady state with respect to a truncated Floquet-Magnus Hamiltonian. 
Using a general argument based on Fermi's golden rule, we show that such Floquet-Gibbs states emerge naturally in periodically modulated Rydberg atomic systems, whose lab-frame Hamiltonian is a quasiperiodic function of time. 
Our approach applies as long as the inherent Bohr frequencies of the system, the modulation frequency and the frequency of the driving laser, which is necessary to uphold high-lying Rydberg excitations, are well separated. 
To corroborate our analytical results, we analyze a realistic model of up to five interacting Rydberg atoms with periodically changing detuning. 
We demonstrate numerically that the second-order Floquet-Gibbs state of this system is essentially indistinguishable from the steady state of the corresponding Redfield equation if the modulation and driving frequencies are sufficiently large. 
\end{abstract}

\maketitle

\newcommand{\ph}{\mathrm{ph}}
\newcommand{\el}{\mathrm{el}}
\newcommand{\rC}{\mathrm{C}}
\newcommand{\rS}{\mathrm{S}}
\newcommand{\rR}{\mathrm{R}}
\newcommand{\tr}{\mathrm{Tr}}
\newcommand{\av}[1]{\langle #1\rangle}
\newcommand{\Hi}{H_\mathrm{int}}
\newcommand{\He}{H_\mathrm{el}}
\newcommand{\oel}{\omega_\mathrm{el}}
\newcommand{\otr}{\omega_\mathrm{tr}}
\newcommand{\od}{\omega_\df}
\newcommand{\kket}[1]{| #1 \rangle \rangle}
\newcommand{\bbra}[1]{\langle \langle #1 |}
\newcommand{\brakket}[1]{\langle #1 \rangle \rangle}
\newcommand{\ttilde}[1]{\tilde{\tilde{ #1}}}
\newcommand{\e}{\mathrm{e}}
\newcommand{\im}{\mathrm{i}}
\newcommand{\df}{\mathrm{d}}

Open quantum systems in weak contact with a thermal environment generically equilibrate to a Gibbs state \cite{ca_1985, isar1994open, car_1999, br_pet_2002, acc_ima_2004, ri_hu_2012, kos_2013}
\begin{equation}
	\varrho_\beta = \e^{-\beta H_\mathrm{S}}/Z_\mathrm{S}.
\end{equation}
Since the partition function $Z_\mathrm{S}$ is determined by the normalization condition $\tr [\varrho_\beta] = 1$, this state depends only on the Hamiltonian $H_\mathrm{S}$ of the system and the inverse temperature $\beta$ of the environment. 
The structure of the system-environment coupling and any dynamical properties of the environment, however, are irrelevant.
Once the system is driven away from equilibrium, such a universal characterization of its state is no longer possible in general \cite{zwan_2001, ma_20006}. 
A rare exception emerges when thermalization is prevented by periodically oscillating fields. 
The Floquet theorem \cite{flo_1883} then guarantees that the stroboscopic dynamics of the isolated system is described by a time-independent effective Hamiltonian $H_\mathrm{F}$ \cite{sam_1973, tor_ku_2005, ki_oka_2011, bu_ale_2015, eck_2017}. 
It thus appears plausible that the system should, at stroboscopic times, settle to a Floquet-Gibbs state of the form 
\begin{equation}
	\varrho_\mathrm{F} = \e^{-\beta H_\mathrm{F}}/Z_\mathrm{F},
\end{equation}
when weakly coupled to a thermal environment \cite{tor_ku_2005, shi_mo_2015, dai_shi_yi_2016, shi_mo_mi_2018, sa_gi_ae_2020, mori_2023}. 
Two objections are commonly raised against this idea \cite{bla_ca_ot_2009, da_ri_2014, ma_char_2016, ku_mo_2016}.
First, the Floquet Hamiltonian $H_\mathrm{F}$ is not unique, since each of its eigenvalues can be arbitrarily shifted by an integer multiple of $\hbar\omega$, where $\omega$ is the driving frequency \cite{mi_pe_1998, tim_rei_1999, kohn_2001}. 
Second, $H_\mathrm{F}$ can usually not be expressed as a sum of local densities, which is a necessary condition for the system to approach a Gibbs-type state \cite{sre_1994, yu_2011, go_eis_2016, da_ka_pol_2016, fa_bran_cra_2017, mo_ike_ka_2018}.  
Both of these problems can be overcome in the high-frequency regime, where the Floquet Hamiltonian can be approximated by a truncation $H_\mathrm{F}^m$ of an asymptotic series, which is known as Floquet-Magnus expansion \cite{bla_ca_ot_2009, ma_2015, mo_ta_2015}. 
This approach fixes eigenvalues of the Floquet Hamiltonian and introduces only a limited degree of non-locality, provided that the original Hamiltonian of the driven system features only short-range interactions and that the truncation order $m$ can be chosen sufficiently small \cite{ku_mo_sai_2016}. 
Indeed, it has been shown that, under certain technical conditions, truncated Floquet-Gibbs state of the form
\begin{equation}\label{eq:TFG}
	\varrho_\mathrm{F}^m = \e^{-\beta H^m_\mathrm{F}}/Z^m_\mathrm{F}
\end{equation}
can, on long intermediate time scales, accurately describe open quantum systems subject to high-frequency periodic driving \cite{ku_mo_sai_2016, shi_ju_2016};
similar conclusions can also be drawn for classical systems, although the Floquet theorem has no strict counterpart for non-linear dynamics \cite{ver_bran_2023_1, ven_kay_2023_2}. 
This phenomenon, which is known as Floquet prethermalization \cite{la_das_moe_2014, wei_knap_2017, wei_knap_2017, mo_2018, her_mu_eck_2018, ru_an_2020, pen_yin_hua_2021, da_den_2021, bea_ja_pi_2021, na_ja_zhou_2022, ho_mo_2023, he_ye_gon_2023}, underpins the idea of Floquet engineering, which aims to create unconventional states of matter by tailoring the effective Hamiltonian of many-body systems with oscillating control fields \cite{bu_ale_2015, ke_de_2018, cla_pan_se_2019, oka_ki_2019, ro_vo_fie_2021, wei_sim_2021, hal_das_2022, cas_de_sa_2022, lu_rei_2022, hao_yue_we_2024}. 

\begin{figure}
\centering
\includegraphics[scale = 0.5]{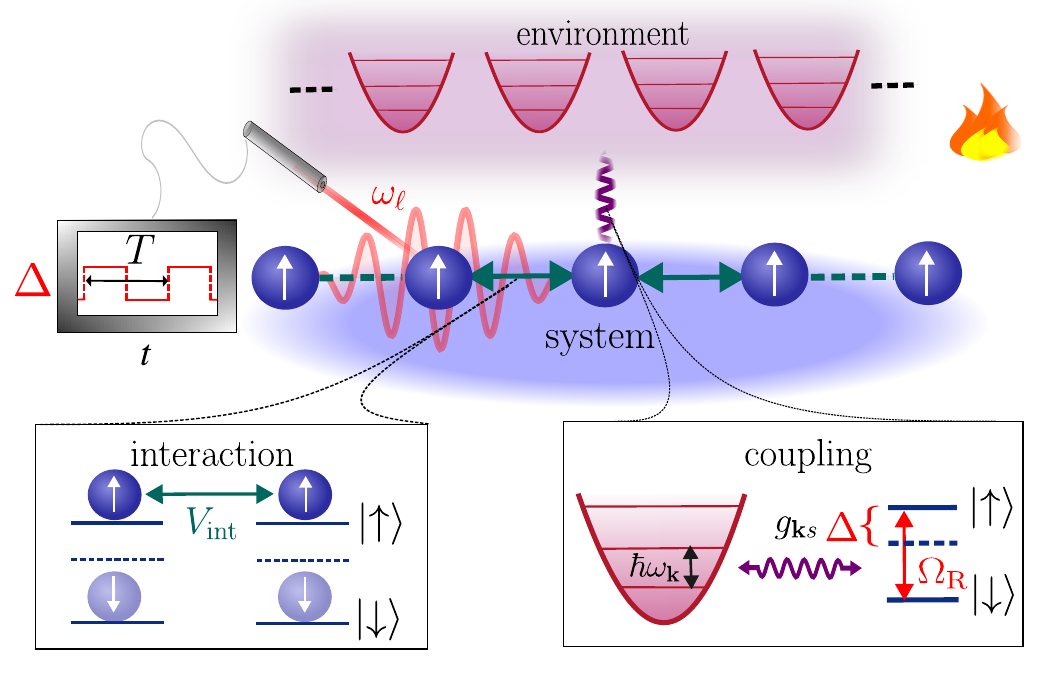}
\caption{\textbf{Sketch of the model.} Our model consists of $L$ Rydberg atoms on a one-dimensional lattice, which are driven by a laser with frequency $\omega_\ell$, detuning $\Delta$ and Rabi frequency $\Omega_\text{R}$. 
The atoms are dipole-coupled to a reservoir of harmonic oscillators describing the modes of a thermal radiation field.  
Nearest-neighbor interactions couple excited atoms with strength $V_\text{int}$. 
A periodic modulation with frequency $\omega=2\pi/T$ is applied to the detuning $\Delta$ by changing the level splitting of the individual atoms, which renders the system Hamiltonian quasiperiodic in time. 
\label{fig:model}}
\end{figure} 

In principle, Rydberg atomic settings offer an excellent platform to apply and test this concept in practice, due to their versatility and high degree of controllability \cite{ga_1988, saf_wal_mo_2010, sch_fel_kol_2011, la_ba_ra_2016, mar_min_2017, tur_mi_2018, lin_mo_2019, adams_prit_2019, lin_ch_hsi_2020, blu_om_le_2021, tur_chri_de_2021}. 
Such systems are shaped by fast oscillating laser fields with frequency $\omega_\ell$, which are required to generate the high-lying excitations that define Rydberg atoms \cite{ci_bla_zo_1992, mu_li_le_2008, noh_ang_2016, eck_2017, bro_thie_2020, ma_nu_wei_2021}.
However, this specific type of driving can usually be eliminated from the system Hamiltonian exactly by means of a local unitary transformation, as is the case for settings, where environmental effects can be neglected  \cite{naz_ni_2024, feld_ni_2024, pi_sa_vi_2024}.
Since $\omega_\ell$ is generally large on any other relevant time scale of the setup, the system thus typically attains a simple Gibbs state in the rotating frame of the driving laser \cite{mori_2023}.
To investigate the emergence of more general Floquet-Gibbs states with Rydberg atoms, it is therefore necessary to introduce an additional periodic modulation with frequency $\omega<\omega_\ell$, see Fig.~\ref{fig:model}.
The situation then becomes more complicated as the system Hamiltonian $H_\mathrm{S} = H_\mathrm{S}(\omega_\ell t,\omega t)$ is now quasiperiodic function of time.
It is thus no longer clear how a truncated Floquet-Gibbs state can be constructed and whether or not it still provides an adequate description of the system on some relevant time scale.

The present article seeks to address this problem. 
As noted above, the laser-induced time dependence of the system Hamiltonian can typically be removed by a local unitary transformation $U(\omega_\ell t) = \bigotimes_j \e^{\im n_j \omega_\ell t}$, where each $n_j$ acts only on a single atom and has no other eigenvalues than $0$ and $\pm 1$ so that $U(\omega_\ell t + 2\pi) = U(\omega_\ell t)\equiv U$. 
In the resulting rotating frame, the joint Hamiltonian of the system and its environment, usually a thermal radiation field, has the general form
\begin{equation}\label{eq:HamFull}
	H'(\omega_\ell t, \omega t) = H'_\mathrm{S}(\omega t) 
			+ \lambda H'_\mathrm{SE}(\omega_\ell t)
			+ H_\mathrm{E}.
\end{equation}
In this expression, $H'_\mathrm{S}(\omega t) = U H_\mathrm{S}(\omega_\ell t, \omega t)U^\dagger - \hbar\omega_\ell \sum_j n_j$ and $H'_\mathrm{SE} = UH_\mathrm{SE}U^\dagger$ are the transformed system and interaction Hamiltonians, $\lambda$ is a small dimensionless coupling parameter and $H_\mathrm{E}$ is the Hamiltonian of the environment. 
Thus, a strictly periodic system Hamiltonian is recovered, but the system-environment coupling now oscillates with the frequency of the laser. 
However, as we show next, this effect is insignificant as long as all relevant time scales are strongly separated. 
Specifically, we require that 
\begin{equation}\label{eq:STS}
	\Omega\ll\omega\ll\omega_\ell, 
\end{equation}
where $\Omega$ is the typical scale of the Bohr frequencies of the average Hamiltonian $H^{\prime 0}_\mathrm{S}= \frac{1}{2\pi}\int_0^{2\pi} \df s \; H'_\mathrm{S}(s)$. 
If these conditions are satisfied, there exists an extended prethermal regime, where the truncated Floquet-Gibbs state \eqref{eq:TFG} accurately describes the stroboscopic state of the driven system in the rotating frame. 
This result constitutes our first main insight and can be understood from the following argument. 

The Floquet-Magnus expansion makes it possible to construct a time-periodic unitary transformation $Q^m(\omega t)\equiv Q^m$ so that $Q^m(2\pi p)=1$ for any integer $p$ and 
\begin{equation}
	Q^m H'_\mathrm{S}(\omega t)Q^{m\dagger} 
		+ i\hbar \dot{Q}^m Q^{m\dagger} 
		= H^{\prime m}_\mathrm{F} + \mathcal{O}(\omega^{-m-1}),
\end{equation}
where the truncated Floquet Hamiltonian $H^{\prime m}_\mathrm{F}$ is time independent \cite{bla_ca_ot_2009}. 
Upon using this transformation to switch to a double-rotating frame, the joint system-environment Hamiltonian \eqref{eq:HamFull} becomes 
\begin{equation}
	H''(t) = H^{\prime m}_\mathrm{F} 
		+ \lambda H^{\prime\prime }_\mathrm{SE}(\omega_\ell t,\omega t) 
		+ H_\mathrm{E} + \mathcal{O}(\omega^{-m-1})
\end{equation}
with $H^{\prime\prime }_\mathrm{SE}(\omega_\ell t,\omega t) = Q^m H'_\mathrm{SE}(\omega_\ell t)Q^{m\dagger}$. 
Since the system Hamiltonian is now time-independent, the problem can, assuming that higher-order corrections in $\omega^{-1}$ are insignificant, be treated with standard methods of time-dependent perturbation theory, where $\lambda$ plays the role of an expansion parameter \cite{merz_1998, me_2014, bal_2014}.  
We denote by $\varepsilon$ and $E$ the eigenvalues of $H^{\prime m}_\text{F}$ and $H_\text{E}$ and by $\ket{\varepsilon}$ and $\ket{E}$ the corresponding eigenstates. 
In lowest order with respect to $\lambda$, the transition rates between the unperturbed states of the joint system are given by Fermi's golden rule, 
\begin{align}\label{eq:FGR}
\Gamma^{\varepsilon'\varepsilon}_{E'E}
	& = \lim_{t\rightarrow\infty}\frac{1}{t}
		\bigl|\bra{\varepsilon', E'}U''(t)\ket{\varepsilon,E}\bigr|^2\\
	& = \frac{2\pi\lambda^2}{\hbar^2}\sum_{q,p}
		\bigl| C^{\varepsilon'\varepsilon}_{E'E}(q,p)\bigr|^2
		\delta\bigl[\Omega_{\varepsilon\varepsilon'}
			+ \Lambda_{EE'} -q \omega_\ell - p\omega\bigr],\nonumber
\end{align}
where $U''(t)$ is the time evolution operator of the joint system in the double-rotating frame. 
Furthermore, $\Omega_{\varepsilon\varepsilon'} = (\varepsilon-\varepsilon')/\hbar$ and $\Lambda_{EE'} = (E - E')/\hbar$ are the Bohr frequencies of the truncated Floquet Hamiltonian and the reservoir Hamiltonian, respectively. 
The sum in Eq.~\eqref{eq:FGR} runs over all integers $q$ and $p$ and the coefficients 
\begin{align}
	& C^{\varepsilon'\varepsilon}_{E'E}(q,p)\\
	& = \frac{1}{4\pi^2}\int_0^{2\pi} \! \df s \int_0^{2\pi}\! \df s' 
		\bra{\varepsilon',E'}H''_\mathrm{SE}(s,s')\ket{\varepsilon,E}
		 \e^{-iqs -ips'}\nonumber
\end{align}
result from a double Fourier expansion of the interaction matrix element. 
It now remains to identify the dominant contributions to the transition rates \eqref{eq:FGR}.

To this end, we note that $|\Omega_{\varepsilon\varepsilon'}|\sim\Omega\ll\omega\ll\omega_\ell$, since $H^{\prime 0}_\text{F}= H^{\prime 0}_\text{S}$ and any higher-order terms in the Floquet-Magnus expansion are parametrically suppressed with inverse powers of $\omega$. 
It is further plausible to assume that $|\Lambda_{EE'}|\sim\Lambda\ll\omega\ll\omega_\ell$ whenever $|C^{\varepsilon'\varepsilon}_{E'E}(q,p)|$ is significant, since any natural interaction Hamiltonian $H_\text{SE}$ should generate environment excitations of the same order of magnitude as the system energy in single-photon processes. 
Thus, for $q\neq 0$, the resonance condition 
\begin{equation}\label{eq:Res}
	\Omega_{\varepsilon\varepsilon'} + \Lambda_{EE'} - q\omega_\ell -p\omega = 0
\end{equation}
can only be met if $|p|\sim (\omega_\ell/\omega)|q| \gg 1$. 
Since, by the Riemann-Lebesgue lemma, the coefficients $C^{\varepsilon'\varepsilon}_{E'E}(q,p)$ decay at least as $1/|p|$ for sufficiently large $|p|$, any such contributions will be subdominant in the sum \eqref{eq:FGR}. 
For $q=0$, however, the condition \eqref{eq:Res} can only be satisfied if $p=0$. 
Hence, up to sub-dominant corrections, the transition rates between the unperturbed states of the joint system are
\begin{equation}
	\Gamma^{\varepsilon'\varepsilon}_{E'E} \simeq \frac{2\pi\lambda^2}{\hbar^2}|C^{\varepsilon'\varepsilon}_{E'E}(0,0)|^2 \delta[\Omega_{\varepsilon\varepsilon'}+\Lambda_{EE'}].
\end{equation}
Upon assuming that the environment is initially in thermal equilibrium at the inverse temperature $\beta$, the transition rates between the free states of the system proper then become 
\begin{equation}
	\Gamma^{\varepsilon'\varepsilon} \simeq \int_0^\infty \! \df E \int_0^\infty \! \df E' \; \rho_{E} \rho_{E'} \cdot 
		\Gamma^{\varepsilon'\varepsilon}_{E'E} e^{-\beta E},
\end{equation}
where $\rho_E$ denotes the density of states of the environment. 
It is now readily verified that these rates satisfy the detailed balance condition
\begin{equation}\label{eq:DB}
	  \Gamma^{\varepsilon'\varepsilon} = \Gamma^{\varepsilon\varepsilon'} e^{-\beta(\varepsilon'-\varepsilon)}
\end{equation}
with respect to the quasi-energies $\varepsilon$ and $\varepsilon'$, i.e., the eigenvalues of the truncated Floquet Hamiltonian $H^{\prime m}_\text{F}$. 
As a result, the system relaxes to a truncated Floquet-Gibbs state of the form \eqref{eq:TFG} in the double-rotating frame, as long as the conditions \eqref{eq:STS} are met. 
In the lab frame, the system thus approaches the quasiperiodic state 
\begin{equation}\label{eq:FGLab}
	\varrho^m_\text{F}(\omega_\ell t, \omega t) = U^\dagger Q^{m\dagger}(e^{-\beta H_\text{F}^m}/Z^m_\text{F})Q^m U,
\end{equation}
which provides a valid description at any time $t$ and is characterized by vanishing net energy absorption. 
This state can further be expected to be stable on an intermediate but practically long time scale, before heating, induced by subdominant corrections to the transition rates, prevails and a trivial infinite temperature state is attained \cite{aba_de_2015, rei_na_2017, ma_ri_2019, tran_eh_2019, mo_2022}.

\begin{figure*}
\centering
\includegraphics[scale = 0.375]{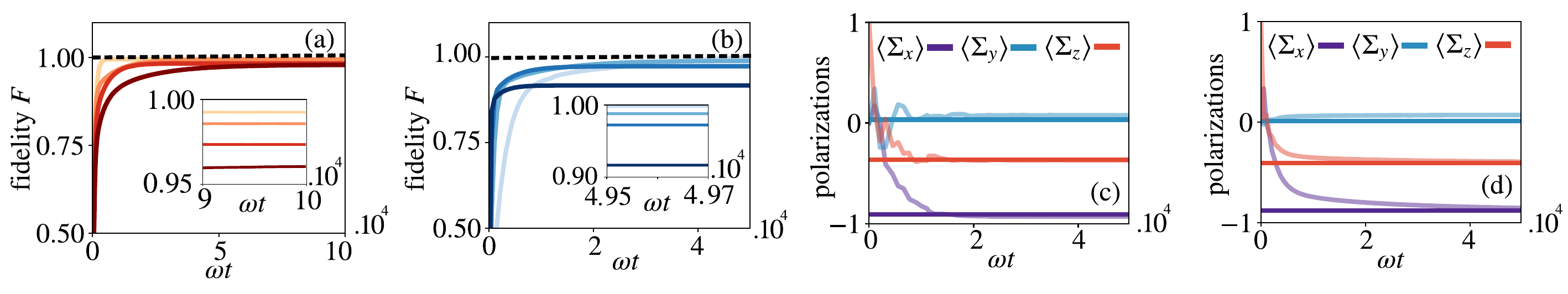}
\caption{\textbf{Quasiperiodic Floquet-Gibbs state in a Rydberg atomic system.}
Panels (a) and (b) show the fidelity between the solution of the Redfield equation \eqref{eq:Red} and the second-order quasiperiodic Floquet-Gibbs state, as defined in Eq.~\eqref{eq:fid}, at stroboscopic times $t_p = 2\pi p/\omega$. 
In (a), we set $V_{\mathrm{int}}/\hbar = \Delta = \Omega_{\mathrm{R}} = 0.1\omega$, and vary the number of atoms $L$ between $2$ to $5$, where darker colors correspond to larger $L$.
In (b) we fix the system size at $L = 3$, and vary the parameters that 
determine the scale $\Omega$ of the Bohr frequencies of the Floquet-Magnus Hamiltonian. 
To this end, we set $V_{\mathrm{int}}/\hbar = \Delta = \Omega_{\mathrm{R}}= 0.1, 0.15, 0.2, 0.25 \omega$, where darker colors correspond to larger frequencies.
Panels (c) and (d) show stroboscopic plots of the mean polarizations $\langle \Sigma_\alpha\rangle = \tr [\Sigma_\alpha \rho^\prime_\text{S}]$, where $\Sigma_\alpha$ is defined in Eq.~\eqref{eq:pol}; horizontal lines show the corresponding expectation values $\langle\Sigma_\alpha\rangle_\text{F} = \tr [\Sigma_\alpha \rho^{\prime 2}_\text{F}]$ in the second-order Floquet-Gibbs ensemble. 
For these plots, we set $V_{\mathrm{int}}/\hbar = \Delta = \Omega_{\mathrm{R}} = 0.1\omega$. 
The system size is $L = 2$ in (c) and $L = 5$ in (d).
For all plots, we have set $\beta = 20/\hbar\omega$ and
$\omega_\ell = 10 \omega$.  
The effective system-environment coupling strength $\kappa$, which absorbs the parameters $g_{\mathbf{k}s}$ and $\omega_\text{k}$ and is defined in Ref.~\cite{supp}, is set to $\kappa = 1/100\pi\omega^{2}$.
The initial state for the Redfield equation \eqref{eq:Red} has been chosen as $\rho^\prime_\text{S}|_{t=0} = \bigotimes_{j} \ket{\uparrow}\bra{\uparrow}$; in Ref. \cite{supp}, we show plots for different initial states.
\label{fig:panel}}
\end{figure*} 

Being reliant on qualitative arguments, the above derivation calls for further confirmation. 
To this end, we consider a generic many-body system of $L$ Rydberg atoms on a one-dimensional lattice, see Fig.~\ref{fig:model}. 
Each atom is described as a two-level system with eigenstates $\ket{\uparrow}$ and $ \ket{\downarrow}$, for each site $j$, and level splitting $\hbar\omega_0$. 
Transitions between the ground and excited states of the individual atoms are driven by an external laser with frequency $\omega_\ell$. 
In the corresponding rotating frame, which is induced by the local unitary transformation $U(\omega_\ell t) = \bigotimes_j \e^{\im n_j \omega_\ell t}$, the system Hamiltonian becomes
\begin{equation}
    H_\mathrm{S}^\prime = \hbar \Delta \sum^{L}_{j = 1} n_j + \hbar\Omega_\text{R} \sum^{L}_{j = 1} \sigma^x_j + H_\mathrm{int}
\end{equation}
with $n = \ket{\uparrow}\bra{\uparrow}$, $\sigma^x =  \ket{\uparrow}  \bra{\downarrow} + \ket{\downarrow}  \bra{\uparrow}$ and $x_j = \mathds{1}^{\otimes(j-1)} \otimes x \otimes \mathds{1}^{\otimes(L-j)}$ throughout.
Furthermore, 
$\Delta = \omega_0-\omega_\ell$ is the detuning of the laser and $\Omega_\text{R}$ the Rabi frequency, which characterizes the coupling strength between laser and atoms. 
The last term, $H_\mathrm{int}$, describes interactions between excited Rydberg atoms, which arise from dipole-dipole or van der Waals forces \cite{wal_saf_2008, mu_li_le_2008,  saf_wal_mo_2010, gam_le_li_2019}. 
For simplicity, we assume these forces to be short-ranged so that they can be modelled with a nearest-neighbor interaction Hamiltonian of the form
\begin{equation}
    H_\mathrm{int} =  V_\mathrm{int} \sum^{L - 1}_{j = 1} n_j n_{j+1},
\end{equation}
where the parameter $V_\mathrm{int}$ sets the interaction strength. 
The environment of the system is formed by a thermal radiation field with bare Hamiltonian 
\begin{equation}\label{bare_res}
    H_\mathrm{E} = \hbar \sum_{\mathbf{k},s} \omega_\mathrm{k}^{\vphantom{\dagger}} a^\dagger_{\mathbf{k}s}a_{\mathbf{k}s}^{\phantom{\dagger}},
\end{equation}
where the Fock-space operators $a_{\mathbf{k}s}^\dagger$ and $a_{\mathbf{k}s}^{\phantom{\dagger}}$ create and annihilate photons with momentum $\mathbf{k}$ and polarization $s$ and $\omega_\mathrm{k} = c\left|\mathbf{k}\right|$ denotes the frequency of the corresponding field mode, with $c$ being the speed of light. 
In the rotating frame of the laser, the general form of the full system-environment Hamiltonian is given by Eq.~\eqref{eq:HamFull}. 
For our model, the system-environment coupling now takes the specific form  
\begin{equation}\label{cou_res}
        H'_\mathrm{SE}(\omega_\ell t) = \hbar \sum_{j, \mathbf{k}, s}
        	\bigl(\e^{\im \omega_\ell t}\sigma_j^+ + \mathrm{h.c.}\bigr)\bigl(g_{\mathbf{k}s}^{\phantom{\dagger}}
        	 a^\dagger_{\mathbf{k}s}  + \mathrm{h.c.}\bigr),
\end{equation}
where $\sigma^+ = \ket{\uparrow}\bra{\downarrow}$ and $\text{h.c}$ stands for Hermitian conjugate.
The coupling constants $g_{\mathbf{k} s}^{\phantom{\dagger}} = \mathbf{d}\cdot \boldsymbol{\epsilon}_s \sqrt{\omega_\mathrm{k}/2\varepsilon_0 \mathcal{V}}$ are determined by the atomic dipole moment $\mathbf{d}$ and the polarization vector $\boldsymbol{\epsilon}_s$ of the corresponding field mode; $\varepsilon_0$ is the dielectric constant of the vacuum and $\mathcal{V}$ the quantization volume \cite{sun_ro_2008, ni_bran_ol_2022}. 

With these prerequisites, we are ready to explore the accuracy of the approximations that lead to the quasiperiodic Floquet-Gibbs state \eqref{eq:FGLab}. 
To introduce a periodic modulation, we assume that the detuning switches between $0$ and some fixed value $\Delta$ at the frequency $\omega = 2\pi/T$. 
That is, we choose the protocol
\begin{equation}
    \begin{aligned}
        \Delta(t) =
        \begin{cases}
         0,  &\text{for} \quad 0 \leq t < T/2 \\
        \Delta, &\text{for} \quad T/2\leq t < T
        \end{cases},
    \end{aligned}
\end{equation} 
where $0\leq t < T$ and $\Delta(t+T) = \Delta(t)$. 
Such a protocol can be realized, for instance, by controlling the level splitting of the individual atoms with external magnetic fields. 
The Floquet Hamiltonian $H'_\text{F}$ can now be calculated perturbatively through the Floquet-Magnus expansion. 
The second-order order truncation of this series, which we use throughout this study, is given by  $H^{\prime 2}_\text{F}= H^{\prime 0}_\text{S} + T H^{\prime (1)}_\text{F} + T^2 H^{\prime (2)}_\text{F}$ with 
\begin{subequations}
\begin{align}
        H^{\prime 0}_\text{S} &= \frac{1}{2} \Bigl( \bar{H}^{\prime}_{\mathrm{S}} + H^{\prime}_{\mathrm{S}}\Bigr),
            \label{eq:FM0}\\
        H^{\prime (1)}_\text{F} &= -\frac{\im}{8\hbar} [ \bar{H}^{\prime}_{\mathrm{S}}, H^{\prime}_{\mathrm{S}}],
            \label{eq:FM1}\\
        H^{\prime (2)}_\text{F} &= -\frac{1}{96\hbar^{2}}\Bigl([H^{\prime}_{\mathrm{S}}, [H^{\prime}_{\mathrm{S}},\bar{H}^{\prime}_{\mathrm{S}}]] + [\bar{H}^{\prime}_{\mathrm{S}}, [ \bar{H}^{\prime}_{\mathrm{S}}, H^{\prime}_{\mathrm{S}}]]\Bigr),
            \label{eq:FM2}
\end{align}
\end{subequations}
and $\bar{H}^{\prime}_\text{S} = H^\prime_\text{S}|_{\Delta =0}$. 
As a benchmark for the corresponding quasiperiodic Floquet-Gibbs state $\varrho^2_\text{F}(\omega_\ell t,\omega t)$, we use the solution of the Redfield equation \cite{red_1965}, which follows from the microscopic model outlined above by tracing out the environmental degrees of freedom and applying the standard Born and Markov approximations. 
These approximations are well justified in the present setting, since the dipole-coupling between atoms and electromagnetic field modes tends to be weak and the thermal radiation field features very short correlation times. 
The Redfield equation, which is given by
\begin{equation}\label{eq:Red}
        \partial_t \varrho'_\text{S} = -\frac{\im}{\hbar}[H_\text{S}^\prime(\omega t), \varrho'_\text{S}] 
        	- [S', G \varrho'_\text{S} - \varrho'_\text{S} G^{\dagger}]
\end{equation}
in the rotating frame of the laser, can therefore be expected to describe our system's actual dynamics accurately \cite{thi_wan_ha_2012, mon_ber_2015, nar_stru_2020, wu_eck_2020, be_wi_2021, da_ca_gi_2023}. 
Here, $\varrho_\text{S}(t)\equiv\varrho_\text{S}$ denotes the evolving density matrix of the system and the time-dependent operators $S'(\omega_\ell t) \equiv S'$ and $G(\omega t)\equiv G$ are given by 
\begin{align}
S' & = \sum_j \e^{\im \omega_\ell t}\sigma_j^+ + \text{h.c.}\;,\\
G & = \int^{\infty}_{0} \df \tau \, C(\tau)U^{\prime}_\text{S}(t) U^{\prime\dagger}_\text{S}(t-\tau) S'(\omega_{\ell} \tau) U^{\prime}_\text{S}(t-\tau) U^{\prime\dagger}_\text{S}(t)
\end{align}
where $U^{\prime\vphantom{\dagger}}_\text{S}(t)$ denotes the time evolution operator of the free system in the rotating frame of the laser, and the bath correlation function is given by 
\begin{equation}
    \label{eq:corr_function}
	C(\tau) = \sum_{\mathbf{k}s} |g_{\mathbf{k}s}^{\vphantom{\dagger}}|^2
		\bigl(\mathfrak{n}_\text{k} \e^{\im \omega_\text{k}\tau} + (1+ \mathfrak{n}_\text{k})\e^{-\im \omega_\text{k}\tau}\bigr)
\end{equation}
with the Bose-Einstein factors $\mathfrak{n}_\text{k}=\bigl[\e^{\beta\hbar\omega_\text{k}}-1\bigr]^{-1}$, for details, see Ref.~\cite{supp}. 

For moderate system size $L$, the Redfield equation \eqref{eq:Red} can be solved numerically as we explain in Ref.~\cite{supp}. 
In Fig.~\ref{fig:panel}(a-d), we compare this solution with the second-order quasiperiodic Gibbs state \eqref{eq:FGLab} for representative parameter values. 
To this end, we calculate the fidelity $F(\varrho'_\text{S}, \varrho^{\prime 2}_\text{F})$, where $\varrho^{\prime 2}_\text{F}(\omega t) = U\varrho^2_\text{F}(\omega_\ell t,\omega t) U^\dagger \equiv \varrho^{\prime 2}_\text{F}$. 
This quantity, which is defined as \cite{niel_chu_2010} 
\begin{equation}
    \label{eq:fid}
        F(X,Y) = \tr\left[\left[
        X^\frac{1}{2}YX^\frac{1}{2}\right]^\frac{1}{2}\right]\leq 1,
\end{equation}
provides a measure for the similarity of two general states $X$ and $Y$ and takes its maximum value $1$ if and only if $X=Y$.  
In addition, we evaluate the expectation values of the three collective polarizations 
\begin{equation}
        \label{eq:pol}
	\Sigma_\alpha = \sum_j \sigma^\alpha_j,
\end{equation}
where $\alpha=x,y,z$ and $\sigma^\alpha$ are the usual Pauli matrices in the basis spanned by the ground and excited states of the individual atoms.
Our numerical results demonstrate that the system, as described by the Redfield equation \eqref{eq:Red}, rapidly approaches a stroboscopic steady state, which is almost indistinguishable from the second-order Floquet-Gibbs state $\varrho^{\prime 2}_\text{F}$, as long as the inherent frequency scale $\Omega$ of the system, the frequency $\omega$ of the periodic modulation and the frequency $\omega_\ell$ of the driving laser are separated by at least one order of magnitude. 
As the separation of time scales is weakened by increasing $\Omega$, subdominant contributions to the transition rates \eqref{eq:FGR} start to play a significant role and deviations between the steady state and the Floquet-Gibbs state become more pronounced.
We further note that the transient period, during which the system approaches its steady state, appears to increase monotonically with the system size $L$, at least for the factorized initial state used in Fig.~\ref{fig:panel}. 
This observation is plausible, since, as the system grows, more and more time will typically be required to build up the long-range correlations encoded in the Floquet-Magnus Hamiltonian; specifically, these correlations are described by the three- and four-body terms that arise from the nested commutators in Eqs. \eqref{eq:FM1} and \eqref{eq:FM2}.
The same behavior is obtained for different initial states, as we have explored in Ref.~\cite{supp}.

In summary, our case study provides ample evidence that periodically modulated Rydberg systems in weak contact with a thermal environment relax to a stable quasiperiodic Floquet-Gibbs state, as long as the three relevant time scales of system, modulation and driving are well separated. 
Notably, this state is stabilized by a strong detuning between the drive and the system-environment coupling, rather than a balance between energy absorption and dissipation, as one might have \emph{a priori} expected. 
The former mechanism leads to vanishing net energy uptake as it effectively insulates the driven system from its environment, once the quasiperiodic state is reached. 
During the transient phase, where the quasiperiodic Floquet-Gibbs state is approached, dissipation is incurred and this process can be described quantitatively with a Lindblad-type master equation, which can be derived from the Redfield equation \eqref{eq:Red} along the same lines that lead to the detailed balance relation \eqref{eq:DB} \cite{da_1974, lin_1976}. 
In the double-rotating frame, this equation takes the form 
\begin{equation}
    \label{eq:lind}
    \partial_t \varrho^{\prime\prime}_\text{S} = -\frac{\im}{\hbar}[H^{\prime m}_\text{F} + H_\text{LS},\varrho^{\prime\prime}_\text{S}] + \mathcal{D}\varrho^{\prime\prime}_\text{S},
\end{equation}
where both the Lamb shift $H_{\mathrm{LS}}$, which commutes with the truncated Floquet-Magnus Hamiltonian $H^{\prime m}_\text{F}$, and the dissipation super operator $\mathcal{D}$ are time-independent; 
the latter has Lindblad form and satisfies the quantum detailed-balance relation  
\begin{equation}
    \mathcal{D}e^{-\beta H^{\prime m}_\text{F}}\circ
    = e^{-\beta H^{\prime m}_\text{F}}\mathcal{D}^\dagger\circ,
\end{equation}
where $\mathcal{D}^\dagger$ denotes the adjoint of $\mathcal{D}$ with respect to the Hilbert-Schmidt scalar product and $\circ$ is a placeholder for an arbitrary operator, for details see Ref.~\cite{supp}.

The general arguments developed in the first part of this article suggest that the behavior seen in our model may in fact be indicative of a more generic phenomenology, which is likely to govern a whole variety of quasiperiodically driven systems in the high-frequency regime. 
We stress that these insights go beyond earlier works, which, to the best of our knowledge, have focused solely on strictly periodic driving. 
Along with experimental tests of our predictions, we leave it to future research to put our theory on rigorous mathematical grounds and to determine its precise range of validity and wider implications \cite{szc_2014}. 
Interesting conceptual directions would be, for example, to characterize the mechanism by which the quasiperiodic state decays once the driving and modulation frequencies fall below certain thresholds or to consider the strong-coupling regime, where Fermi's golden rule no longer applies. 
On the more practical side, further exploring how Rydberg atomic systems may be used to realize unconventional states of matter or how quasiperiodic driving can be deployed in quantum thermodynamics to engineer new types of thermal machines provide compelling topics for further investigations \cite{szc_gel_ali_2013, car_gam_bran_2020, mar_ca_li_2023}.

\textbf{Data access statement:} The numerical data supporting the findings of this article and the code used to generate them are available on Zenodo \cite{wilson_2024_floquet_gibbs}.

\begin{acknowledgments}
\textbf{Acknowledgments:}
We acknowledge funding from the Deutsche Forschungsgemeinschaft (DFG, German Research Foundation) through the Research Unit FOR 5413/1, Grant No. 465199066. This project has also received funding from the European Union’s Horizon Europe research and innovation program under Grant Agreement No. 101046968 (BRISQ). F.C.~is indebted to the Baden-W\"urttemberg Stiftung for the financial support by the Eliteprogramme for Postdocs. This work was supported by the University of Nottingham and the University of T\"{u}bingen’s funding as part of the Excellence Strategy of the German Federal and State Governments, in close collaboration with the University of Nottingham. 
This work was supported by the Medical Research Council (Grants No. MR/S034714/1 and MR/Y003845/1) and the Engineering and Physical Sciences Research Council (Grant No. EP/V031201/1).
\end{acknowledgments}

\bibliography{main}

\end{document}


\preprint{APS/123-QED}

\newcommand{\ph}{\mathrm{ph}}
\newcommand{\el}{\mathrm{el}}
\newcommand{\rC}{\mathrm{C}}
\newcommand{\rS}{\mathrm{S}}
\newcommand{\rR}{\mathrm{R}}
\newcommand{\tr}{\mathrm{Tr}}
\newcommand{\av}[1]{\langle #1\rangle}
\newcommand{\Hi}{H_\mathrm{int}}
\newcommand{\He}{H_\mathrm{el}}
\newcommand{\oel}{\omega_\mathrm{el}}
\newcommand{\otr}{\omega_\mathrm{tr}}
\newcommand{\od}{\omega_\df}
\newcommand{\kket}[1]{| #1 \rangle \rangle}
\newcommand{\bbra}[1]{\langle \langle #1 |}
\newcommand{\brakket}[1]{\langle #1 \rangle \rangle}
\newcommand{\ttilde}[1]{\tilde{\tilde{ #1}}}
\newcommand{\e}{\mathrm{e}}
\newcommand{\im}{\mathrm{i}}
\newcommand{\df}{\mathrm{d}}

\newpage

\renewcommand\thesection{S\arabic{section}}
\renewcommand\theequation{S\arabic{equation}}
\renewcommand\thefigure{S\arabic{figure}}
\setcounter{equation}{0}
\setcounter{figure}{0}

\onecolumngrid

\newpage

\setcounter{page}{1}
\widetext
\begin{center}
{\Large SUPPLEMENTAL MATERIAL}
\end{center}
\begin{center}
\vspace{0.8cm}
{\Large {Quasiperiodic Floquet-Gibbs states in Rydberg atomic systems}}
\end{center}

\begin{center}
Wilson S. Martins$^1$, Federico Carollo$^2$, Kay Brandner$^{3,4}$ and Igor Lesanovsky$^{1,3,4}$ 
\end{center}
\begin{center}
$^1${\em Institut f\"{u}r Theoretische Physik and Center for Integrated Quantum Science and Technology,}\\
{\em  Universit\"{a}t T\"{u}bingen, Auf der Morgenstelle 14, 72076 T\"{u}bingen, Germany,} \\
$^2$ {\em Centre for Fluid and Complex Systems, Coventry University, Coventry, CV1 2TT, United Kingdom}\\
$^3${\em School of Physics and Astronomy, University of Nottingham, Nottingham, NG7 2RD, UK}\\
$^4${\em Centre for the Mathematics and Theoretical Physics of Quantum Non-Equilibrium Systems,}\\
{\em University of Nottingham, Nottingham, NG7 2RD, UK}\\

\end{center}

In the first two sections of this Supplemental Material, we derive the Redfield and Floquet-Lindblad equations, which were introduced Eqs.\,(\textcolor{blue}{21}) and (\textcolor{blue}{27}) of the main text. 
In the third section, we provide details on the numerical calculations that delivered the results shown in Fig.\,\textcolor{blue}{2} of the main text. 

\section{Redfield equation}

The Redfield equation (\textcolor{blue}{21}) can be derived through standard techniques, which we briefly recapitulate here for the sake of completeness \cite{red_1965}. 
Our starting point is the joint Hamiltonian of system and environment in the rotating frame of the driving laser,
\begin{equation}
    H'(t) = H'_\mathrm{S}(\omega t) + \lambda H'_\mathrm{SE}(\omega_\ell t) + H_\mathrm{E},
\end{equation}
where $H'_\text{S}(\omega t+2\pi) = H'_\text{S}(\omega t)$ and $H'_\text{SE}(\omega_\ell t+2\pi) = H'_\text{SE}(\omega_\ell t) = S'(\omega_\ell t) \otimes E$ with $S'(\omega_\ell t)=S^{\prime\dagger}(\omega_\ell t)$ and $E=E^\dagger$ being the coupling operators of the system and the environment. 
Upon making the ansatz $\varrho'(t) = \varrho'_\mathrm{S}(t) \otimes \pi_\beta + \chi(t)$ for the joint density matrix, where $\pi_\beta = e^{-\beta H_\text{E}}/Z_\text{E}$ is the thermal state of the environment and the traceless correction $\chi(t)$ accounts for system-environment correlations, the Liouville–von Neumann equation can be decomposed into a system of two coupled differential equations, 
\begin{subequations}
\begin{align}
    \label{eq:LES}
    \im \hbar \partial_t\varrho'_\mathrm{S}(t) & = [H'_\mathrm{S}(\omega t), \varrho'_\mathrm{S}(t)] + \lambda \tr_\mathrm{E}\bigl[[H'_\mathrm{SE}(\omega_\ell t), \chi(t)]\bigr],\\
    \label{eq:LEC}
    \im\hbar \partial_t \chi(t) & = [H'_\mathrm{S}(\omega t) + H_\mathrm{E}, \chi(t)] + \lambda [H'_\mathrm{SE}(\omega_\ell t), \chi(t)] + \lambda [H'_\mathrm{SE}(\omega_\ell t), \varrho'_\mathrm{S}(t) \otimes \pi_\beta] - \lambda \tr_\mathrm{E}\bigl[[H'_\mathrm{SE}(\omega_\ell t), \chi(t)]\bigr]\otimes \pi_\beta.
\end{align}
\end{subequations}
Here, we assume, without loss of generality, that $\tr_\text{E}[E \pi_\beta]=0$, where $\tr_\text{E}[\cdots]$ denotes the partial trace over environmental degrees of freedom. 
Solving Eq.~\eqref{eq:LEC} perturbatively in $\lambda$ to first order for the initial condition $\chi(0) = 0$, i.e., $\rho'(0)=\rho'_\text{S}(0)\otimes \pi_\beta$, and inserting the result into Eq.~\eqref{eq:LES} yields the Born equation 
\begin{equation}
    \label{eq:born_eq}
     \partial_t\tilde{\varrho}'_{\mathrm{S}}(t) = - \int^{t}_{0} \df \tau \, C(t - \tau) [\tilde{S}'(\omega_{\ell}t, t), \tilde{S}'(\omega_{\ell} \tau, \tau) \tilde{\varrho}'_{\mathrm{S}}(\tau)] + \text{h.c.} \;,
\end{equation}
where $\text{h.c.}$ stands for the Hermitian conjugate.
Tildes indicate the interaction picture with respect to the bare system evolution, i.e.,
\begin{equation}
    \tilde{\varrho}'_\text{S}(t)=U^{\prime\dagger}_\text{S}(t)\varrho'_\text{S}(t)U'_\text{S}(t) 
    \quad\text{and}\quad
    \tilde{S}'(\omega_{\ell} t, t) = U^{\prime\dagger}_\text{S}(t) S'(\omega_{\ell} t) U^{\prime}_\text{S}(t),
\end{equation}
where $U'_\text{S}(t)$ satisfies $\im\hbar\partial_t U'_\text{S}(t) = H'_\text{S}(\omega t)U'_\text{S}(t)$ and $U'_\text{S}(0)=1$. 
The environment correlation function is defined as 
\begin{equation}
    C(t) = \frac{\lambda^2}{\hbar^2} \tr[E(t) E \pi_\beta]
    \quad\text{with}\quad
    E(t) = \e^{\im H_{\mathrm{E}}t/\hbar} E \e^{-\im H_{\mathrm{E}}t/\hbar}.
\end{equation}
To eliminate memory effects from Eq.~\eqref{eq:born_eq}, we apply the standard Markov approximation, which yields the result
\begin{equation}
    \label{eq:SMRedTrans}
    \partial_t\tilde{\varrho}'_{\mathrm{S}}(t) = - \int^{t}_{-\infty} \df \tau \, C(t - \tau) [\tilde{S}'(\omega_{\ell}t, t), \tilde{S}'(\omega_{\ell} \tau,  \tau) \tilde{\varrho}'_{\mathrm{S}}(t)] + \text{h.c.} \;.
\end{equation}
Upon undoing the transformation to the interaction picture, we thus obtain the Redfield equation 
\begin{equation}
    \label{eq:SMRed}
          \partial_t\varrho'_{\mathrm{S}} =  -\frac{\im}{\hbar} [H'_{\mathrm{S}}(\omega t) \varrho'_{\mathrm{S}}] - [S'(\omega_{\ell}t), G(\omega t) \varrho'_{\mathrm{S}}(t)] + \text{h.c.}\;,
\end{equation}
where we have defined the operator 
\begin{align}
     G(\omega t) & = \int^{t}_{-\infty} \df \tau \, C(t - \tau)U^{\prime}_\text{S}(t) U^{\prime\dagger}_\text{S}(\tau) S'(\omega_{\ell} \tau) U^{\prime}_\text{S}(\tau) U^{\prime\dagger}_\text{S}(t)\\
     & = \int_0^\infty \df \tau \, C(\tau) Q(\omega t)e^{-\im H'_\text{F}\tau/\hbar}Q^\dagger(\omega t- \omega\tau)
     S'(\omega_\ell \tau) Q(\omega t - \omega\tau) e^{\im H'_\text{F}\tau/\hbar}Q^\dagger(\omega t)\nonumber
\end{align}
with $U'_\text{S}(t) = Q(\omega t) e^{-\im H'_\text{F} t/\hbar}$ being the Floquet decomposition of the time evolution operator of the bare system. 

\section{Floquet-Lindblad equation}
\newcommand{\ve}{\varepsilon}
To derive the Floquet-Lindblad equation (\textcolor{blue}{27}), we first recall that, for sufficiently large modulation frequencies $\omega$, the time evolution operator of the system in the rotating frame of the laser can be approximated by the  $m^\text{th}$-order truncation of the Floquet-Magnus expansion, $U'_\text{S}(t) \simeq Q^m(\omega t) e^{-\im H^{\prime m}_\text{F}t/\hbar}$.
With this approximation, we have 
\begin{equation}
    \tilde{S}'(\omega_\ell t, t) = 
    e^{\im H^m_\text{F} t/\hbar}Q^{m\dagger}(\omega t) 
    S'(\omega_\ell t) Q^m(\omega t) e^{-\im H^m_\text{F}t/\hbar}
    = \sum_{\ve\ve'}
    \ket{\ve}\bra{\ve'}
    S''_{\ve\ve'}(\omega_\ell t,\omega t)
    e^{\im\Omega_{\ve\ve'} t}
\end{equation}
with $\ve$ and $\ket{\ve}$ being the eigenvalues and eigenstates of $H^{\prime m}_\text{F}$ and  $\Omega_{\ve\ve'} = (\ve-\ve')/\hbar$ the corresponding Bohr frequencies; $S''_{\ve\ve'}(\omega_\ell t,\omega t) = \bra{\ve} Q^{m\dagger}(\omega t)S(\omega_\ell t)Q^m(\omega t)\ket{\ve'}$ denotes the matrix element of the system coupling operator in the double-rotating frame.
The latter object can be decomposed in a double Fourier series, 
\begin{equation}
    S''_{\varepsilon\varepsilon'}(\omega_\ell t,\omega t) 
        = \sum_{qp} S''_{\varepsilon\varepsilon'}(q,p)
        e^{\im q\omega_\ell + \im p\omega t}
        \quad\text{with}\quad
    S''_{\varepsilon\varepsilon'}(q,p) 
    = \frac{1}{4\pi^2}\int_0^{2\pi} \! ds \int_0^{2\pi} \! ds'
        \; S''_{\varepsilon\varepsilon'}(s,s') e^{-\im sq -\im ps'}, 
\end{equation}
where $q$ and $p$ run over all integers. 
Inserting this expansion into \eqref{eq:SMRedTrans} and using that the operator $\tilde{S}(\omega_\ell t,\omega t)$ is self-adjoint, yields 
\begin{equation}\label{eq:SMRedExp}
    \partial_t \tilde{\varrho}'_\text{S}(t) = -\sum_{\ve\ve'}\sum_{\bar{\ve}\bar{\ve}'}
    \sum_{qp}\sum_{\bar{q}\bar{p}}
    e^{-\im [(q-\bar{q})\omega_\ell + (p-\bar{p})\omega 
        +(\Omega_{\ve\ve'}-\Omega_{\bar{\ve}\bar{\ve}'})]t}
    \tilde{C}(\bar{q}\omega_\ell + \bar{p}\omega + \Omega_{\bar{\ve}\bar{\ve}'})
    S^{\prime\prime\ast}_{\ve\ve'}(q,p)
    S^{\prime\prime}_{\bar{\ve}\bar{\ve}'}(\bar{q},\bar{p})
    [\Pi_{\ve'\ve},\Pi_{\bar{\ve}\bar{\ve}'}
    \tilde{\varrho}'_\text{S}(t)]
    +\text{h.c.}
\end{equation}
with 
\begin{equation}
    \Pi_{\ve\ve'}\equiv\ket{\ve}\bra{\ve'}
    \quad\text{and}\quad
    \tilde{C}(\upsilon) \equiv \int^{\infty}_{0} \df \tau \,C(\tau) e^{-\im\upsilon \tau}
\end{equation}
being the one-sided Fourier transform of the environment correlation function. 
We now apply the rotating-wave approximation, where we follow the arguments outlined in the main text. 
We first observe that the phase in Eq.~\eqref{eq:SMRedExp} consists of three terms, which are proportional to $\omega_\ell$, $\omega$ and $|\Omega_{\ve\ve'}|\sim\Omega$, respectively. 
Provided that all time scales are well separated, these terms cannot compensate each other without suppressing the Fourier coefficients $S''_{\ve\ve'}(q,p)$, which have to decay at least as $|S''_{\ve\ve'}(q,p)|\sim |q|^{-1}|p|^{-1}$ according to the Riemann-Lebesgue lemma. 
Hence, all contributions to the sum in Eq.~\eqref{eq:SMRedExp}, for which $q\neq \bar{q}$, $p\neq \bar{p}$ or $\Omega_{\ve\ve'}\neq \Omega_{\bar{\ve}\bar{\ve}'}$ will either be subdominant or average out. 
Upon keeping only the most relevant contributions, we obtain the master equation 
\begin{equation}\label{eq:SMFLindbladPrim}
    \partial_t \tilde{\varrho}'_\text{S}(t) = 
    - \sum_{qp}\sum_\Omega \tilde{C}(q\omega_\ell + p\omega + \Omega)[L^\dagger_\Omega(q,p),L_\Omega(q,p)\tilde{\varrho}'_\text{S}(t)] + \text{h.c.}
    \quad\text{with}\quad
    L_{\Omega}(q,p) \equiv \sum_{\ve\ve'\; : \; \Omega_{\ve\ve'}=\Omega}
    S''_{\ve\ve'}(q,p)\Pi_{\ve\ve'},
\end{equation}
which can be shown to have Lindblad form. 
To this end, we decompose the one-sided Fourier transform of the environment correlation function into its real and imaginary part, 
\begin{equation}
    \label{eq:SM_FTCF}
    \tilde{C}(\upsilon) = \frac{\gamma(\upsilon)}{2} + \im \varphi(\upsilon),
    \quad\text{where} \quad
    \gamma(\upsilon) = \tilde{C}(\upsilon)
    + \tilde{C}^\ast(\upsilon) = \int_{-\infty}^\infty \df\tau \; 
    C(\tau) e^{-i \upsilon \tau},
\end{equation}
since $C^\ast(\tau) = C(-\tau)$. 
With this decomposition, Eq.~\eqref{eq:SMFLindbladPrim} becomes
\begin{equation}\label{eq:SMFLindbladSecnd}
    \partial_t \tilde{\varrho}'_\text{S}(t)= 
    - \frac{\im}{\hbar}[H_\text{LS},\tilde{\varrho}'_\text{S}(t)]
    + \sum_{qp}\sum_{\Omega}\frac{\gamma(q\omega_\ell + p\omega + \Omega)}{2}
    \left(
        L_\Omega(q,p)\tilde{\varrho}'_\text{S}(t)L^\dagger_\Omega(q,p)
        -L^\dagger_\Omega(q,p)L_\Omega(q,p)\tilde{\varrho}'_\text{S}(t)
        +\text{h.c.}
    \right),
\end{equation}
where we have introduced the Lamb shift Hamiltonian 
\begin{equation}
    H_\text{LS}= H^\dagger_\text{LS} 
        = \hbar\sum_{qp}\sum_\Omega\varphi(q\omega_\ell + p\omega +\Omega) L^\dagger_\Omega(q,p) L_\Omega(q,p).
\end{equation}
Since $C(\tau)$ can be shown to be a function of positive type, its Fourier transform $\gamma(\upsilon)$ must be non-negative by Bochner's theorem \cite{li_2020}. 
Hence, the master equation \eqref{eq:SMFLindbladSecnd} does indeed have Lindblad form. 
Furthermore, the Lamb shift and the Lindblad operators satisfy the commutation relations 
\begin{equation}
    [H^{\prime m}_\text{F},H_\text{LS}] = 0,
    \quad
    [H^{\prime m}_\text{F},L_\Omega(q,p)] = \hbar\Omega L_\Omega(q,p),
    \quad
    [H^{\prime m}_\text{F},L_\Omega^\dagger (q,p)] = - \hbar\Omega L_\Omega^\dagger (q,p).
\end{equation}
Therefore, we can switch to the double-rotating frame through the transformation $\varrho_\text{S}''(t) = e^{-\im H^{\prime m}_\text{F} t/\hbar} \tilde{\varrho}'_\text{S}(t) e^{\im H^{\prime m}_\text{F} t/\hbar}$ without making the left-hand side of Eq.~\eqref{eq:SMFLindbladSecnd} time dependent. 
That is, we have 
\begin{equation}\label{eq:SMFLindbladThird}
    \partial_t \varrho''_\text{S}(t)= 
    - \frac{\im}{\hbar}[H^{\prime m}_\text{F} + H_\text{LS},\varrho''_\text{S}(t)]
    + \sum_{qp}\sum_{\Omega}\frac{\gamma(q\omega_\ell + p\omega + \Omega)}{2}
    \left(
        L_\Omega(q,p)\varrho''_\text{S}(t)L^\dagger_\Omega(q,p)
        -L^\dagger_\Omega(q,p)L_\Omega(q,p)\varrho''_\text{S}(t)
        +\text{h.c.}
    \right). 
\end{equation}
This equation can be further simplified if the environment correlation function $C(\tau)$ decays on a characteristic time scale $\tau_\text{rel}$ that is short compared to the typical system-environment interaction time, which is set by the coupling parameter $\lambda$, and long compared to the periods $2\pi/\omega_\ell$ and $2\pi/\omega$ of the laser driving and the periodic modulation. 
Under this condition, which is analogous to the assumption $|\Lambda_{EE'}|\sim\Lambda\ll \omega \ll \omega_\ell$ used in the main text, the rate function $\gamma(\upsilon)$ decays at least as $\gamma(\upsilon)\sim (\upsilon\tau_\text{rel})^{-1}$, as can be shown with the help of the Riemann-Lebesgue lemma. 
Hence, the dominant contribution to the sum over Fourier modes in Eq.~\eqref{eq:SMFLindbladThird} comes from the term $p=q=0$. 
Upon neglecting subdominant corrections, we end up with the reduced master equation 
\begin{equation}\label{eq:SMFLindbladRed}
    \partial_t \varrho''_\text{S}(t) = - \frac{\im}{\hbar}[H^{\prime m}_\text{F} + H_\text{LS},\varrho''_\text{S}(t)]
    + \mathcal{D}\varrho''_\text{S}(t),
\end{equation}
where we have introduced the super operator 
\begin{equation}
    \mathcal{D}\circ \equiv \sum_\Omega
    \gamma(\Omega)\left(
    L_\Omega^{\vphantom{\dagger}} \circ L_\Omega^\dagger - \frac{1}{2}L^\dagger_\Omega L_\Omega^{\vphantom{\dagger}}\circ 
    -\frac{1}{2}\circ L^\dagger_\Omega L_\Omega^{\vphantom{\dagger}}
    \right)
    \quad\text{with}\quad
    L_\Omega \equiv L_\Omega(q=0,p=0). 
\end{equation}
Since $\gamma(\Omega) = e^{-\beta \hbar \Omega}\gamma(-\Omega)$ \cite{li_2020}, it is now straightforward to verify that the dissipation super operator satisfies the quantum detailed-balance relation 
\begin{equation}
    \mathcal{D}e^{-\beta H^{\prime m}_\text{F}}\circ = 
    e^{-\beta H^{\prime m}_\text{F}}\mathcal{D}^\dagger \circ 
    \quad\text{with}\quad
    \mathcal{D}^\dagger \circ \equiv 
    \sum_\Omega
    \gamma(\Omega)\left(
    L_\Omega^{\dagger} \circ L_\Omega^{\vphantom{\dagger}} - \frac{1}{2}L^\dagger_\Omega L_\Omega^{\vphantom{\dagger}}\circ 
    -\frac{1}{2}\circ L^\dagger_\Omega L_\Omega^{\vphantom{\dagger}}
    \right).
\end{equation}
Hence, provided that the set of Lindblad operators $\{L_\Omega^{\vphantom{\dagger}},L^\dagger_\Omega\}$ is irreducible, the unique stationary state of Eq.~\eqref{eq:SMFLindbladRed} is given by the Floquet-Gibbs state $\rho^{\prime\prime m}_\text{F} = e^{-\beta H^{\prime m}_\text{F}}/Z^m_\text{F}$ \cite{da_1974, li_2020}.

\section{Simulating the Redfield equation}

The simulations presented in this work are based on quantities defined in the rotating frame of the laser; the time-dependent system Hamiltonian $H'_{\text{S}}=H'_\text{S}(\omega t)$, the coupling operator $S' = S'(\omega_{\ell} t)$ and the one-sided Fourier transform of the environment correlation function $\tilde{C}(\upsilon)$.
These quantities are sufficient for solving the Redfield equation using the Bloch-Redfield tensor, implemented by the QuTiP toolbox, as detailed in Refs.~\cite{joh_2012, joh_na_no_2013}.
The form of the function $\tilde{C}(\upsilon)$ can be derived by taking the continuum limit of the modes of the thermal radiation field, which make up the system's environment. 
Upon applying standard techniques \cite{br_pet_2002}, this procedure yields the result 
\begin{equation}
    \tilde{C}(\upsilon) = \int^\infty_0 \text{d}\tau \int^{\infty}_0 \text{d}\omega_\text{k} \omega^3_\text{k}\Bigl[\kappa e^{-\im (\omega_\text{k} - \upsilon)\tau}\bigl[1 + \mathfrak{n}_\text{k}\bigr] + \kappa e^{\im (\omega_\text{k} + \upsilon) \tau}\mathfrak{n}_\text{k}\Bigr],
\end{equation}
where $\kappa$ sets the overall time scale of the system-environment interactions and absorbs the coupling constants $g_{\mathbf{k}s}$ and the mode frequencies $\omega_\text{k}$ in the continuum limit. 
Hence, the decay rate $\gamma (\upsilon) = \tilde{C}(\upsilon) + \tilde{C}^\ast(\upsilon)$ becomes $\gamma(\upsilon) = 2\pi \kappa \upsilon^{3} \bigl[1 + \mathfrak{n}(\upsilon)\bigr]$ and 
the function $\varphi(\upsilon)$, which determines the Lamb-shift Hamiltonian defined in Eq.~\eqref{eq:SM_FTCF}, is given by 
\begin{equation}
    \varphi(\upsilon) = \kappa  \int^{\infty}_0 \mathrm{d}\omega_\mathrm{k} \omega^3_\mathrm{k} \left[P\left(\frac{1 + \mathfrak{n}_\text{k}}{\upsilon - \omega_\text{k}} \right) + P\left(\frac{\mathfrak{n}_\text{k}}{\upsilon + \omega_\text{k}} \right)\right].
\end{equation} 
Here, $\mathfrak{n}(\upsilon)=\bigl[\e^{\beta\hbar\upsilon}-1\bigr]^{-1}$ are Bose-Einstein factors satisfying $\mathfrak{n}(-\upsilon) = - \bigl[1 + \mathfrak{n}(\upsilon)\bigr]$ and $P(\circ)$ represents the Cauchy principal value.

\begin{figure*}
\centering
\includegraphics[scale = 0.45]{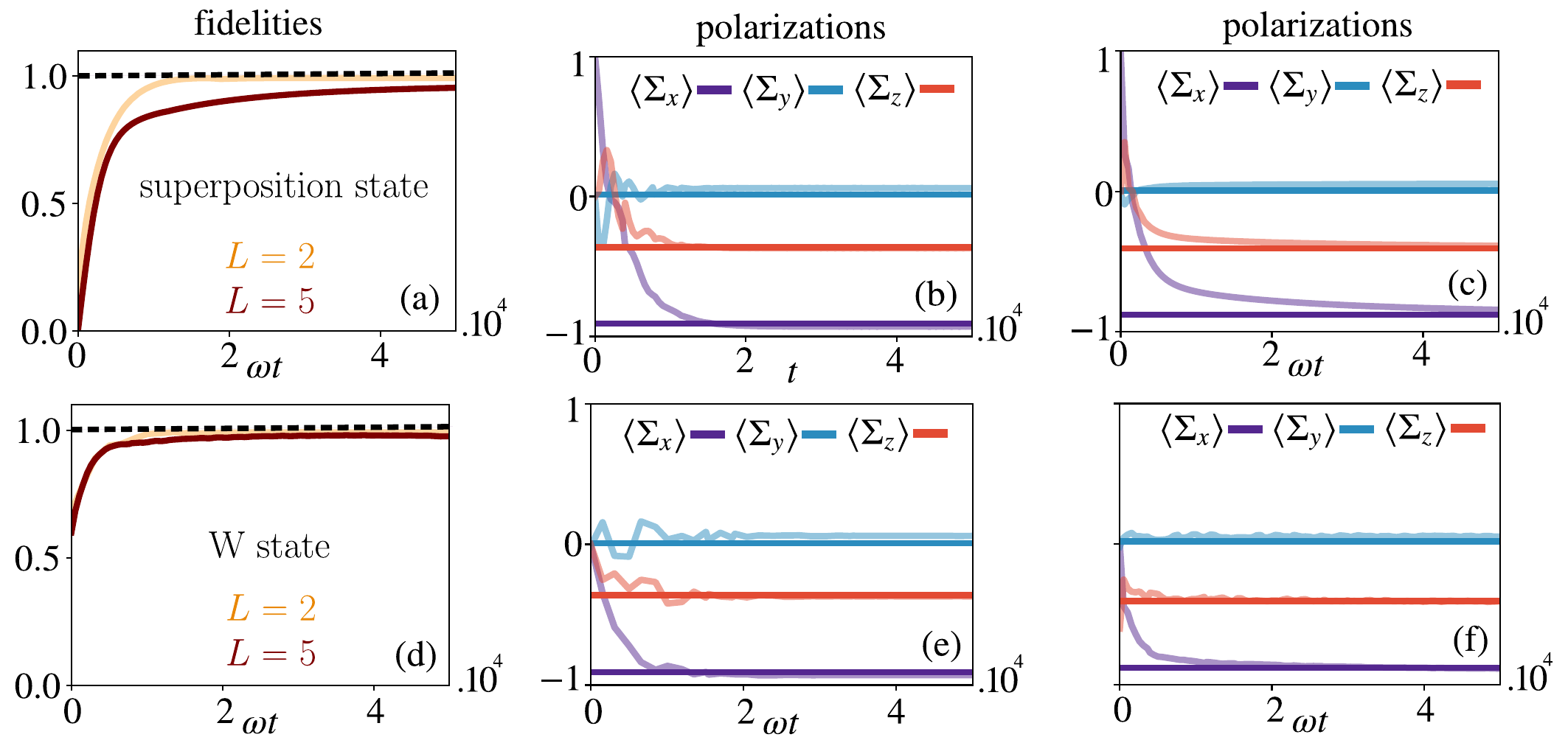}
\caption{\textbf{Redfield equation vs quasiperiodic Floquet-Gibbs state for different initial states.}
We compare the solutions of the Redfield equation (\textcolor{blue}{21}) for different initial states with the second-order Floquet-Gibbs $\varrho^{\prime 2}_\text{F}$. 
The initial state is chosen as $\rho^\prime_\text{S}|_{t=0} = \ket{\psi_{0}} \bra{\psi_{0}}$, where $\ket{\psi_{0}}$ is given by Eqs.~\eqref{eq:ssstate} and \eqref{eq:wstate} for the first and second row, respectively.
Figures (a) and (d) show the fidelities between the corresponding solution of the Redefield equation, (\textcolor{blue}{21}), and the second-order Floquet Gibbs state, $\varrho^{\prime 2}_\text{F}$, as defined in Eq.~(\textcolor{blue}{25}) of the main text, for $L = 2$ (yellow line) and $L = 5$ (red line). 
Figures (b) and (c) and (e) and (f) show the corresponding polarizations for $L = 2$ and $L = 5$, respectively.
In theses plots, horizontal lines indicate the expectation values $\langle\Sigma_\alpha\rangle_\text{F} = \tr [\Sigma_\alpha \rho^{\prime 2}_\text{F}]$ in the second-order Floquet-Gibbs ensemble.  
For all plots, we have set $V_{\mathrm{int}}/\hbar = \Delta = \Omega_{\mathrm{R}} = 0.1\omega$, $\beta = 20/\hbar\omega$ and
$\omega_\ell = 10 \omega$.  
The effective system-environment coupling strength is set to $\kappa = 1/100\pi\omega^{2}$.
\label{fig:sup_panel}}
\end{figure*}

In the main text, we compare the results of our numerical solution of the Redfield equation (\textcolor{blue}{21}) with those obtained from the truncated Floquet-Gibbs state, which is defined in Eq.\,(\textcolor{blue}{3}) of the main text.
This comparison is also made in Fig. \ref{fig:sup_panel} for different initial states.
In Figs. \ref{fig:sup_panel}(a--c) we use the factorized superposition state 
\begin{equation}\label{eq:ssstate}
    \ket{\psi_0} = \bigotimes^{L}_{j = 1}\frac{\ket{\downarrow} + \ket{\uparrow}}{\sqrt{2}},
\end{equation}
and, in Figs. \ref{fig:sup_panel}(d--f), we use the symmetric W state
\begin{equation}\label{eq:wstate}
    \ket{\psi_0} = \frac{\ket{\uparrow \downarrow \downarrow ... \downarrow} + \ket{\downarrow \uparrow \downarrow ... \downarrow} + ... \ket{\downarrow \downarrow  ... \downarrow \uparrow}}{\sqrt{L}},
\end{equation}
defining a initial density matrix $\rho^\prime_\text{S}|_{t=0} = \ket{\psi_{0}} \bra{\psi_{0}}$.
The behavior shown in the plots indicates that the decay modes associated with W states give rise to shorter transient periods before approaching the steady state as compared to product states \cite{bel_gi_pal_2017}. 
For these results and those of the main text, we neglect the contribution from the Lamb-Shift Hamiltonian term \cite{co_gla_2023, ma_ro_2023}.
Our annotated code and the data used in this study are publicly available on Zenodo \cite{wilson_2024_floquet_gibbs}.

\bibliography{supp.bib}